\documentclass[onecolumn,showpacs,superscriptaddress,prd]{revtex4}
\usepackage{amsmath}
\usepackage{amsfonts}
\usepackage{amssymb}
\usepackage{graphicx}
\usepackage{hyperref}
\usepackage{units}
\usepackage{color}
\usepackage{ulem}

\def\b{\begin{equation}}
\def\e{\end{equation}}
\def\ba{\begin{eqnarray}}
\def\ea{\end{eqnarray}}

\def\u{\uparrow}
\def\d{\downarrow}
\def\la{\langle}
\def\ra{\rangle}

\def\re{\rangle}

\begin{document}

\title{Entanglement in  anisotropic expanding spacetime}

\author{Roberto Pierini\footnote{robpierin@gmail.com}}
\affiliation{Institute of Theoretical Physics, University of Warsaw, Pasteura 5, 02-093 Warsaw, Poland}

\author{Shahpoor Moradi\footnote{moradis@ucalgary.ca}}
\affiliation{University of Calgary, Department of Geoscience, Calgary, Canada}

\author{Stefano Mancini\footnote{email: stefano.mancini@unicam.it}}
\affiliation{School of Science and Technology, University of Camerino, I-62032 Camerino, Italy}
\affiliation{INFN-Sezione di Perugia, I-06123 Perugia, Italy}

\date{\today}

\begin{abstract}
We study the effect of space anisotropy in the entanglement generated by expanding universe on spin $0$ and $\frac{1}{2}$ fields.
For massive scalar field we find revivals of entanglement entropy vs momentum
after decreasing from the maximum at $k=0$.
In massive Dirac field the effect is a slight distortion of the non-monotonic profile giving rise to the maximum of entanglement entropy at $k> 0$.
More interestingly, massless field of both type can only get entangled through anisotropy,
with a maximum of entanglement entropy occurring at $k> 0$.
\end{abstract}

%\keywords{Quantum fields in curved spacetime, Entanglement characterization}

\maketitle

\section{Introduction}

The advent of quantum theory of information stimulated investigations across traditional frontiers. Among others we are witnessing an extension of quantum information notions into the general relativity framework.
As a consequence entanglement, being recognized as a fundamental resource in quantum information processing, has been widely investigated in curved spacetime. 
In particular it has been realized the possibility of generating it
by the expansion of the universe \cite{Ball06}, \cite{MPM14} and \cite{MM12}.
This effect can be traced back to the mechanism of particle-antiparticle production during cosmic evolution, a phenomenon pointed out time ago \cite{Parker}.

The vast majority of studies along this line focus on homogeneous and isotropic spacetime, that is using a Friedman-Robertson-Walker (FRW) background \cite{BD01}.
However, our universe is neither homogeneous nor isotropic: there are structures in it, galaxies, clusters of galaxies, superclusters, etc. which causes deviations from the FRW background.
Even in the very early universe one expects quantum fluctuations from this FRW background to occur.
Then, an interesting question is if we could in principle observe, by means of particle correlations, any departure from homogeneity and isotropy.

Thus we want, at least partially, release the symmetry assumptions for the metric and
consider anisotropy, i.e. a universe whose expansion rate depends on the direction. There is a currently going on debate regarding the possible deviation from isotropy of the nowadays expansion rate of the universe, see \cite{Con1,Con4,Con2} for observational constraints.
Here, we refer to a model introduced in \cite{ZS01}, where the authors resorted to the case of small changes in the metric, in order to study general properties of particles production. Its cosmological importance resides on the fact that such small changes in the metric can have an observable effect on the anisotropy of the cosmic microwave background radiation. Spacetime metrics whose scaling factors along the three spatial axes depend on time in a different way are not conformally flat, and as such they can lead to strong particles production independently from the mass. After having specified the vacuum state of the matter field (propagating over the given spacetime background) in the remote past, one can study all the properties of the excited states in the far future, when the process is completed. We are especially interested in the quantum correlations between the created particles. However,
dealing with anisotropic spacetime models present technical difficulties and solutions can only be pursued by resorting to perturbative approaches \cite{Lot01,Lot02}.

Resorting to these techniques, entanglement for the scalar field has been recently studied by us in the anisotropic spacetime context \cite{PMM17}.
Here, we enlarge this study to encompass also the Dirac field and provide a comparison between the two cases (spin-less and spin-$\frac{1}{2}$ field).
We found marked differences between scalar and Dirac particles for the massive case
and similarities for the massless case.
In fact, while for massive scalar field revivals of entanglement entropy vs momentum appears after decreasing from the maximum at $k = 0$, in the massive Dirac field we find just a slight distortion of the non-monotonic profile giving rise to the maximum of entanglement entropy at
$k> 0$. In contrast, it turns out that massless {particles} of both type can only get entangled through anisotropy, with a maximum of entanglement entropy occurring at $k> 0$.

The paper is structured as follows. In Sec. \ref{section2} we review the mechanism that leads to entanglement in expanding universe under the assumption of homogeneous and isotropic spacetime. Then, in Sec. \ref{section3} we present the model of anisotropic spacetime and its solution. The results about entanglement following from it are discussed in Sec. \ref{section4}. Sec. \ref{section5} is for final remarks.

%%%%%%%%%%%%%%%%%%%%%%%%%%%%%%%%%%%%%%%%%%%%%%%%%%%%%%%%

\section{Entanglement in homogeneous and isotropic spacetime}\label{section2}

We start revisiting the mechanism that leads to entanglement in expanding universe under the assumption of homogeneous spacetime, i.e. by considering a FWR metric \cite{BD01}
\begin{equation}
ds^2=a^2(\eta)\left\{ d\eta^2-(dx^1)^2-(dx^2)^2-(dx^3)^2\right\},
\end{equation}
where $a(\eta)$ is the scale factor with $\eta$ the conformal time.

\subsection{Scalar field}

Given a scalar field $\phi(\vec x,t)$ we assume that it can be decomposed through two different set of orthonormal modes $\{\phi^{in}_i,\phi_i^{in*}\}$ and $\{\phi_i^{out},\phi_i^{out*}\}$. The labels $in$ and $out$, refer to two different Minkowskian spacetime regions. Physically, $|0\re_{in}$
is the state with no incoming particles (anti-particles) in the remote
past and $|0\re_{out}$ is the state with no outgoing
particles (anti-particles) in the far future. In those regions the spacetime is flat and
the dynamics of the field is that of the free field. So we have two natural modes decompositions
of the field, associated with two Fock spaces.
Elements in the basis set are orthonormal, which means $(\phi_i,\phi_j)=\delta_{ij}$, $(\phi_i^*,\phi_j^*)=-\delta_{ji}$ and $(\phi_i,\phi_j^*)=0$, according to the Klein-Gordon inner product.

Following the standard procedure, we associate to each mode $\phi^{in/out}_{\vec k}$ and to its complex conjugate $\phi_{\vec k}^{in/out*}$ annihilation and creation operators, $ a_{\vec k}^{in/out}$ and $a^{in/out\,\dagger}_{\vec k}$, satisfying the equal time commutation relations
$\left[ a_{\vec k}^{in/out}\,, a^{in/out\,\dagger}_{\vec k'}\right]=\delta_{\vec k\vec k'}$ and $\left[ a_{\vec k}^{in/out}\,, a^{in/out}_{\vec k'}\right]=0$.
The two set of modes define then two representations of the scalar field
\begin{subequations}\label{scadec}
\begin{align}
\phi&=\sum_{\vec k}\left\{ a^{in}_{\vec k}\phi^{in}_{\vec k}+a^{in\,\dagger}_{\vec k}\phi^{in*}_{\vec k}\right\}\\
&=\sum_{\vec k'}\left\{a_{\vec k'}^{out}\phi^{out}_{\vec k'}+a^{out\,\dagger}_{k'}\phi_{\vec k}^{out*}\right\}  \,.
\end{align}
\end{subequations}
Expanding one mode in terms of the others
\begin{equation}
\phi^{out}_{\vec k'}=\sum_{\vec k}\left\{ \alpha_{\vec k'\vec k}\phi^{in}_{\vec k}+\beta_{\vec k'\vec k}\phi^{in*}_{\vec k}\right\} \,,
\end{equation}
and inserting it in (\ref{scadec}) gives a map between $in$ and $out$ operators
\begin{subequations}\label{dir-bog-tra}
\begin{align}
&a_{\vec k}^{in}=\sum_{\vec k'}\left\{ \alpha_{\vec k'\vec k}a^{out}_{\vec k'}+\beta_{\vec k'\vec k}^{*} a^{\dagger\,out}_{\vec k'}\right\} \,,  \\
&a^{out}_{\vec k'}=\sum_{\vec k}\left\{ \alpha^*_{\vec k'\vec k} a^{in}_{\vec k}-\beta^*_{\vec k'\vec k}a^{\dagger\,in}_{\vec k}\right\}\,.
\end{align}
\end{subequations}
The coefficients $\alpha$ and $\beta$ are known as Bogolubov coefficients and are defined as $\alpha_{ij}=(\phi^{out}_i,\phi_j^{in})$ and $\beta_{ij}=-(\phi^{out}_i,\phi_j^{in*})$.
Inserting Eq.(\ref{dir-bog-tra}) into the equal time commutation relations yields
\begin{subequations}\label{bog-relations}
\begin{align}
&\sum_{\vec k}\left( \alpha_{\vec k_1\vec k}^*\alpha_{\vec k_2\vec k}-\beta^*_{\vec k_1\vec k}\beta_{\vec k_2\vec k}  \right)=\delta_{\vec k_1\,,\vec k_2},\\
&\sum_{\vec k}\left( \alpha_{\vec k_1\vec k}\beta_{\vec k_2\vec k}-\beta_{k_1k}\alpha_{k_2k}  \right)=0   \,.
\end{align}
\end{subequations}
A great simplification comes from homogeneity of spacetime, for which the metric does not have any space dependence. The consequence is that the Fourier decomposition of the field results in plane waves propagating throughout the spacetime, each with proper wave lenght $\phi_k(x)=\chi_k(t)e^{i\vec k\cdot\vec x}$. The modes are then decoupled and the Bogolubov coefficients can be written as
\begin{subequations}
\begin{align}
\alpha_{\vec k'\vec k}&=\alpha_{\vec k} \,\delta_{\vec k\,,\vec k'}, \\
\beta_{\vec k'\vec k}&=\beta_{\vec k} \,\delta_{\vec k\,,-\vec k'} \,.
\end{align}
\end{subequations}
Plugging these into (\ref{dir-bog-tra}) gives diagonal Bogolubov transformations
\begin{subequations}\label{inv-bog-tra1}
\begin{align}
&a^{in}_{\vec k}=\alpha_{\vec k} a^{out}_{\vec k}+\beta_{\vec k}^* a_{-\vec k}^{\dagger\,out} \,,  \\
&a^{out}_{\vec k}=\alpha^*_{\vec k} a^{in}_{\vec k}-\beta_{\vec k}^* a_{-\vec k}^{\dagger\,in} \,,
\end{align}
\end{subequations}
where the mixing is only between modes of opposite momentum.
Moreover, Eq.(\ref{bog-relations}) simplifies in
\b\label{norm1}
|\alpha_{\vec k}|^2-|\beta_{\vec k}|^2=1   \,.
\e
The most important consequence of the coupling implied by Eq. (\ref{dir-bog-tra}) is easily shown.
Suppose the field is in the vacuum state of the $in$ modes and we want to evaluate the expectation value of the particle number operator for the $out$ modes. All we have to do is to insert the second equation $(\ref{inv-bog-tra1})$ and its complex conjugate into ${}_{in}\langle0|a_{\vec k}^{\dagger\,out}a_{\vec k}^{out}|0\rangle_{in}$. The result is
\b\label{part-number}
{}_{in}\langle0|a_{\vec k}^{\dagger\,out}a_{\vec k}^{out}|0\rangle_{in}
=|\beta_{\vec k}|^2 \,.
\e
Therefore, the vacuum input state is not empty in the $out$ region. In the context of cosmology
$|\beta_{\vec k}|^2$ is interpreted as the number of created particles per mode due to the expansion of the universe \cite{Parker}. These particles come in pair of opposite momentum and they lead to entanglement.

If we consider the $in$-vacuum as a pure state of a bi-partite system labeled by the momenta $\vec k$ and $-\vec k$, it can be decomposed as
\b\label{sch-dec}
|0_{\vec k}\,0_{-\vec k}\ra^{in}=\sum_{n=0}^{\infty}c_n \;|n\ra_{\vec k}^{out}|n\ra_{-\vec k}^{out} \,.
\e
The coefficients $c_n$ in (\ref{sch-dec}) are evaluated through $a^{in}_{\vec k}|0\ra^{in}=0$ and Eq.(\ref{inv-bog-tra1}). The result is
\begin{align*}
0=&(\alpha_{\vec k}a_{\vec k}^{out}+\beta_{\vec k}^* a_{-\vec k}^{\dagger\,out}) \sum_{n=0}^{\infty}c_n \;|n\ra_{\vec k}^{out}|n\ra_{-\vec k}^{out}   \\
=&\sum_{n=1}^{\infty}\alpha_{\vec k}\,c_n \sqrt n\;|n-1\ra_{\vec k}^{out}|n\ra_{-\vec k}^{out}+\sum_{n=0}^{\infty}\beta_{\vec k}^*\,c_n\sqrt{n+1} \;| n\ra_{\vec k}^{out}|n+1\ra_{-\vec k}^{out} \\
=&\sum_{n=1}^{\infty}(\alpha_{\vec k}\,c_{n+1}+\beta_{\vec k}^*\,c_n\sqrt{n+1})\;|{n}\ra_{\vec k}^{out}|n+1\ra_{-\vec k}^{out}  \,.
\end{align*}
Being this a linear superposition of independent vectors, it must be
$ \alpha_{\vec k}\,c_{n+1}+\beta_{\vec k}^*\,c_n=0 $, and then
\b
c_n=\left(-\frac{\beta_{\vec k}^*}{\alpha_{\vec k}}\right)^n \, c_0 \,.
\e
Furthermore, imposing the normalization condition we get $c_0=\sqrt{1-\left|\frac{\beta_{\vec k}^*}{\alpha_{\vec k}}\right|^2}$.

To determine the entanglement content of the state $\varrho_{\vec k,-\vec k}^{out}$ from (\ref{sch-dec}),
 we derive the reduced density operator by tracing over antiparticles
\begin{align}\label{redscal}
\varrho_{\vec k}^{(out)}={\rm Tr}_{-\vec k}\left(\varrho_{\vec k,-\vec k}^{(out)}\right)
=\left(1-\left|\frac{\beta_{\vec k}^*}{\alpha_{\vec k}}\right|^2\right)\sum_{n=0}^{\infty}
\left|\frac{\beta_{\vec k}^*}{\alpha_{\vec k}}\right|^{2n} |n\rangle_{\vec k}\langle n|^{out},
\end{align}
and then evaluate its von Neumann entropy
\b
S(\vec k)=-{\rm Tr} \left(\varrho^{out}_{\vec k}\log_2 \varrho^{out}_{\vec k}\right).
\e

%%%%%%%%%%%%%%%%%%%%%%%%%%%%%%%%%%%%%%%%%%%%%%%%%%%%%%%

\subsection{Dirac field}

Dirac field describes fermions whose spin is $\frac{1}{2}$.
The inner product for half spin particles is positive definite, while the Klein-Gordon inner product is not,  $(\psi_1,\psi_2)^{\dagger}=(\psi_1,\psi_2)^*=(\psi_2,\psi_1)$.
As for the scalar field the Dirac field also can be decomposed in $in$-modes and $out$-modes. Assuming we have two complete basis $\{ \psi_{\vec k\,\sigma}^{+\,in},\psi_{\vec k\,\sigma}^{-\,in} \}$, 
$\{ \psi_{\vec k\,\sigma}^{+\,out},\psi_{\vec k,\sigma}^{-\,out} \}$, where $\psi_{\vec k\,\sigma}^{\pm}$ refers to positive/negative frequency mode of momentum $\vec k$ and spin $\sigma$, which define
particles and anti-particles in asymptotic regions and have corresponding
vacua $|0\re_{in}$ and $|0\re_{out}$. Also in case of Dirac particles we assume that the modes decompositions is valid in the $in$ and $out$ regions. Hence, we can write
\begin{subequations}
\begin{align}
\label{Psi-dec-in}
\psi=&\sum_{\vec k'\sigma'}\left( a_{\vec k'\,\sigma'}^{in}\psi_{\vec k'\,\sigma'}^{+\,in}+b_{\vec k'\,\sigma'}^{\dagger\,in}\psi_{\vec k'\,\sigma'}^{-\,in} \right) \\
\label{Psi-dec-out}
=&\sum_{\vec k\sigma}\left(a_{\vec k\,\sigma}^{out}\psi_{\vec k\,\sigma}^{+\,out}+b_{\vec k\,\sigma}^{\dagger\,out}\psi_{\vec k\,\sigma}^{-\,out} \right)  \,,
\end{align}
\end{subequations}
where now creation and annihilation operators satisfy the anti-commutation relations
\begin{subequations}
\begin{align}
&\left\{a_{\vec k\,\sigma}^{in/out}\,,a_{\vec k'\,\sigma'}^{\dagger\,in/out}\right\}=\delta_{\vec k,\vec k'}\delta_{\sigma,\sigma'} \,, \\
&\left\{a_{\vec k\,\sigma}^{in/out}\,,a_{\vec k'\,\sigma'}^{in/out}\right\}=\left\{a_{\vec k\,\sigma}^{\dagger\,in/out}\,,a_{\vec k'\,\sigma'}^{\dagger\,in/out}\right\}=0   \,.
\label{algf}
\end{align}
\end{subequations}
Likewise the scalar field, it is possible to relate the operators of
$in$-particles to those of $out$-particles through Bogolubov transformations. Namely, we can write
\begin{subequations}\label{psi-in-out}
\begin{align}
\psi^{+\,in}_{\vec k'\,\sigma'}=&\sum_{\vec k\sigma}\left[ \alpha_{\vec k'\vec k\sigma'\sigma}\,\psi_{\vec k\,\sigma}^{+\,out}+\beta_{\vec k'\vec k\sigma'\sigma}\,\psi_{\vec k\,\sigma}^{-\,out} \right]\,, \\
\psi^{-\,in}_{\vec k'\,\sigma'}=&\sum_{\vec k\sigma}\left[\xi_{\vec k'\vec k\sigma'\sigma}\,\psi_{\vec k\,\sigma}^{+\,out}+\zeta_{\vec k'\vec k\sigma'\sigma}\,\psi_{\vec k\,\sigma}^{-\,out} \right]   \,,
\end{align}
\end{subequations}
with the Bogolubov coefficients defined as
\begin{subequations}\label{dir-bog-coef}
\begin{align}
\alpha_{\vec k'\vec k\sigma'\sigma}&=\left(\psi_{\vec k\,\sigma}^{+\,out},\psi^{+\,in}_{\vec k'\,\sigma'}\right) \,, \qquad\qquad
\xi_{\vec k'\vec k\sigma'\sigma}=\left(\psi_{\vec k\,\sigma}^{+\,out},\psi^{-\,in}_{\vec k'\,\sigma'}\right)   \,, \\
\beta_{\vec k'\vec k\sigma'\sigma}&=\left(\psi_{\vec k\,\sigma}^{-\,out},\psi^{+\,in}_{\vec k'\,\sigma'}\right) \,,  \qquad\qquad
\zeta_{\vec k'\vec k\sigma'\sigma}=\left(\psi_{\vec k\,\sigma}^{-\,out},\psi^{-\,in}_{\vec k'\,\sigma'}\right)   \,.
\end{align}
\end{subequations}
Plugging Eq. (\ref{psi-in-out}) into (\ref{Psi-dec-in}) and comparing with (\ref{Psi-dec-out}), straightforwardly gives
\begin{subequations}\label{dir-bog}
\begin{align}
&a_{\vec k\,\sigma}^{out}=\sum_{\vec k'\sigma'}\big[ \alpha_{\vec k'\vec k\sigma'\sigma}\,a_{\vec k'\,\sigma'}^{in}+\xi_{\vec k'\vec k\sigma'\sigma}\,b^{\dagger \, in}_{\vec k'\,\sigma'} \big] \,,  \\
&b_{\vec k\,\sigma}^{\dagger\,out}=\sum_{\vec k'\sigma'}\big[\beta_{\vec k'\vec k\sigma'\sigma}\,a_{\vec k'\,\sigma'}^{in}+ \zeta_{\vec k'\vec k\sigma'\sigma}\,b^{\dagger \, in}_{\vec k'\,\sigma'} \big] \,,
\end{align}
\end{subequations}
and the inverse
\begin{subequations}
\begin{align}
a_{\vec k'\,\sigma'}^{in}&=\sum_{\vec k\sigma}\big[\alpha^*_{\vec k'\vec k\sigma'\sigma}\,a_{\vec k\,\sigma}^{out}+ \beta^*_{\vec k'\vec k\sigma'\sigma}\,b^{\dagger\,out}_{\vec k\,\sigma} \big] \,,  \\
b_{\vec k'\,\sigma'}^{\dagger\,in}&=\sum_{\vec k\sigma}\big[\zeta^*_{\vec k'\vec k\sigma'\sigma}\,b_{\vec k\,\sigma}^{\dagger\,out}+\xi^*_{\vec k'\vec k\sigma'\sigma}\,a^{out}_{\vec k\,\sigma} \big] \,.
\end{align}
\end{subequations}
In homogeneous spacetime also the Dirac field propagates as a plane wave
\b
\psi_{\vec k\,\sigma}^{+\,in/out}(\vec x,t)=\psi_{\vec k\,\sigma}^{+\,in/out}(t)\,e^{i\vec k\cdot\vec x} \,,
\e
which implies
\begin{subequations}
\begin{align}
\alpha_{\vec k\vec k'\sigma\sigma'}&=\alpha_{\vec k\sigma\sigma'}\delta_{\vec k\,,\vec k'}\,, \qquad\qquad
\xi_{\vec k\vec k'\sigma\sigma'}=\xi_{\vec k\sigma\sigma'}\delta_{\vec k\,,-\vec k'}\,,  \\
\beta_{\vec k\vec k'\sigma\sigma'}&=\beta_{\vec k\sigma\sigma'}\delta_{\vec k\,,-\vec k'}\,,  \qquad\qquad
\zeta_{\vec k\vec k'\sigma\sigma'}=\zeta_{\vec k\sigma\sigma'}\delta_{\vec k\,,\vec k'}   \,.
\end{align}
\end{subequations}
That gives us the following form for the Bogolubov transformations
\begin{subequations}\label{dir-bog}
\begin{align}
a_{\vec k\,\sigma}^{in}&=\sum_{\sigma'}\big[ \alpha^*_{\vec k\sigma\sigma'}a_{\vec k\,\sigma'}^{out}+ \beta^*_{\vec k\sigma\sigma'}b^{\dagger\,out}_{-\vec k\,\sigma'} \big]  \,,  \\
b_{\vec k\,\sigma}^{\dagger\,in}&=\sum_{\sigma'}\big[\zeta^*_{\vec k\sigma\sigma'}\,b_{\vec k\,\sigma'}^{\dagger\,out}+\xi^*_{\vec k\sigma\sigma'}\,a^{out}_{-\vec k\,\sigma'} \big] \,.
\end{align}
\end{subequations}
with coefficients satisfying
\b
\sum_{\sigma}[{\alpha_{\vec k\,\sigma'\sigma}}^*\alpha_{\vec k\,\sigma''\sigma}+{\beta_{\vec k\,\sigma'\sigma}^*}\beta_{\vec k\,\sigma''\sigma}]=\delta_{\vec k\,\sigma'\sigma''} \,.
\e
Because of no modes mixing we are allowed to focus on a single mode $k$ and the $in$-vacuum state associated to it can be written as a linear superposition of the $out$-basis states
\begin{align}
|0_{\vec k};0_{-\vec k}\ra^{in}=&c_1|0_{\vec k};0_{-\vec k}\ra^{out}
+c_2|\u_{\vec k};\d_{-\vec k}\ra^{out}
+c_3|\d_{\vec k};\u_{-k}\ra^{out}  \\
+&c_4|\u_{\vec k};\u_{-k}\ra^{out}+c_5|\d_{\vec k};\d_{-\vec k}\ra^{out}
+c_6|\u\d_{\vec k},\u\d_{-\vec k}\ra^{out} \,.
\end{align}
This explicit form is subjected to Pauli exclusion principle and to charge super-selection rule.
Now, proceeding as in the scalar case, imposing $a_{k,\sigma}^{in}|0_{\vec k};0_{-\vec k}\ra^{in}=0$, we get
\begin{align}
|0_{\vec k};0_{-\vec k}\ra^{in}=N\Big[&
H_{00}|0_{\vec k};0_{-\vec k}\ra^{out}+H_{\u\d} |\u_{\vec k};\d_{-\vec k}\ra^{out}
+H_{\d\u}|\d_{\vec k};\u_{-k}\ra^{out}  \nonumber \\
&+H_{\u\u}|\u_{\vec k};\u_{-k}\ra^{out}+H_{\d\d}|\d_{\vec k};\d_{-\vec k}\ra^{out} +H_{33}|\u\d_{\vec k},\u\d_{-\vec k}\ra^{out}\Big] \,.
\label{inout}
\end{align}
where
\begin{align}
\begin{split}
H_{00}&=\alpha^*_{\u\d}\alpha^*_{\d\u}-\alpha^*_{\u\u}\alpha^*_{\d\d}\,,  \\
H_{\u\d}&=\alpha^*_{\d\d}\beta^*_{\u\d}-\alpha^*_{\u\d}\beta^*_{\d\d}\,,  \\
H_{\d\u}&=\alpha^*_{\u\u}\beta^*_{\d\u}-\beta^*_{\u\u}\alpha^*_{\d\u}\,,  \\
H_{\u\u}&=\alpha^*_{\d\d}\beta^*_{\u\u}-\alpha^*_{\u\d}\beta^*_{\d\u}\,,  \\
H_{\d\d}&=\alpha^*_{\u\u}\beta^*_{\d\d}-\beta^*_{\u\d}\alpha^*_{\d\u}\,,  \\
H_{33}&=\beta^*_{\u\u}\beta^*_{\d\d}-\beta^*_{\u\d}\beta^*_{\d\u}  \,.
\end{split}
\label{h}
\end{align}
and $|N|^2=[|H_{00}|^2+|H_{\u\d}|^2+|H_{\d\u}|^2+|H_{\u\u}|^2+|H_{\d\d}|^2+|H_{33}|^2]^{-1}$.
It is well known that any Bogolubov transformation can be represented by a unitary operator
mapping $in$ into $out$ ladder operators. The same unitary also realizes the transformation of $in$ to $out$ states. This means that the transformation (\ref{inout}) is given by a unitary $\mathcal U$ corresponding to equation (\ref{dir-bog}). The operator $\mathcal U$ is discussed explicitly in Ref. \cite{PMM16}.

The reduced density operator for particles is obtained by tracing over antiparticles the state  $\varrho_{\vec k,-\vec k}^{(out)}$ as \eqref{inout} and it results in
\begin{align}
\varrho_{\vec k}^{(out)}&={\rm Tr}_{-\vec k}\left(\varrho_{\vec k,-\vec k}^{(out)}\right)\\
&= |N|^2
\Big[|H_{00}|^2|0_{\vec k}\ra\la 0_{\vec k}|
+(|H_{\u\d}|^2+|H_{\u\u}|^2) |\u_{\vec k}\ra\la\u_{\vec k}| +(|H_{\d\u}|^2+|H_{\d\d}|^2) |\d_{\vec k}\ra\la\d_{\vec k}|
+ |H_{33}|^2 |\u\d_{\vec k}\ra\la\u\d_{\vec k}|   \nonumber\\
&\hspace{1.3cm}+(H_{\u\d}^*H_{\d\d}+H_{\u\u}^*H_{\d\u})|\d_{\vec k}\ra\la\u_k| +(H_{\d\d}^*H_{\u\d}+H_{\u\u}H_{\d\u}^*)|\u_{\vec k}\ra\la\d_k|\Big] \,,
\label{reducedrho}
\end{align}

Then, also in this case the entanglement entropy is evaluated as
\begin{align}\label{Dir-entr-funct}
S(\vec k)&=-{\rm Tr}\left(\varrho_{\vec k}^{(out)}\log_2 \varrho_{\vec k}^{(out)}\right).
\end{align}

%%%%%%%%%%%%%%%%%%%%%%%%%%%%%%%%%%%%%%%%%%%%%%%%%%%%%%%%%%%%%%%%%%%%%%%%%%%%%%%%%%%%%%%%%%%%%%%%%%%%%%

\section{Anisotropic spacetime model}\label{section3}

\begin{figure}
\centering
\includegraphics[width=6.5in]{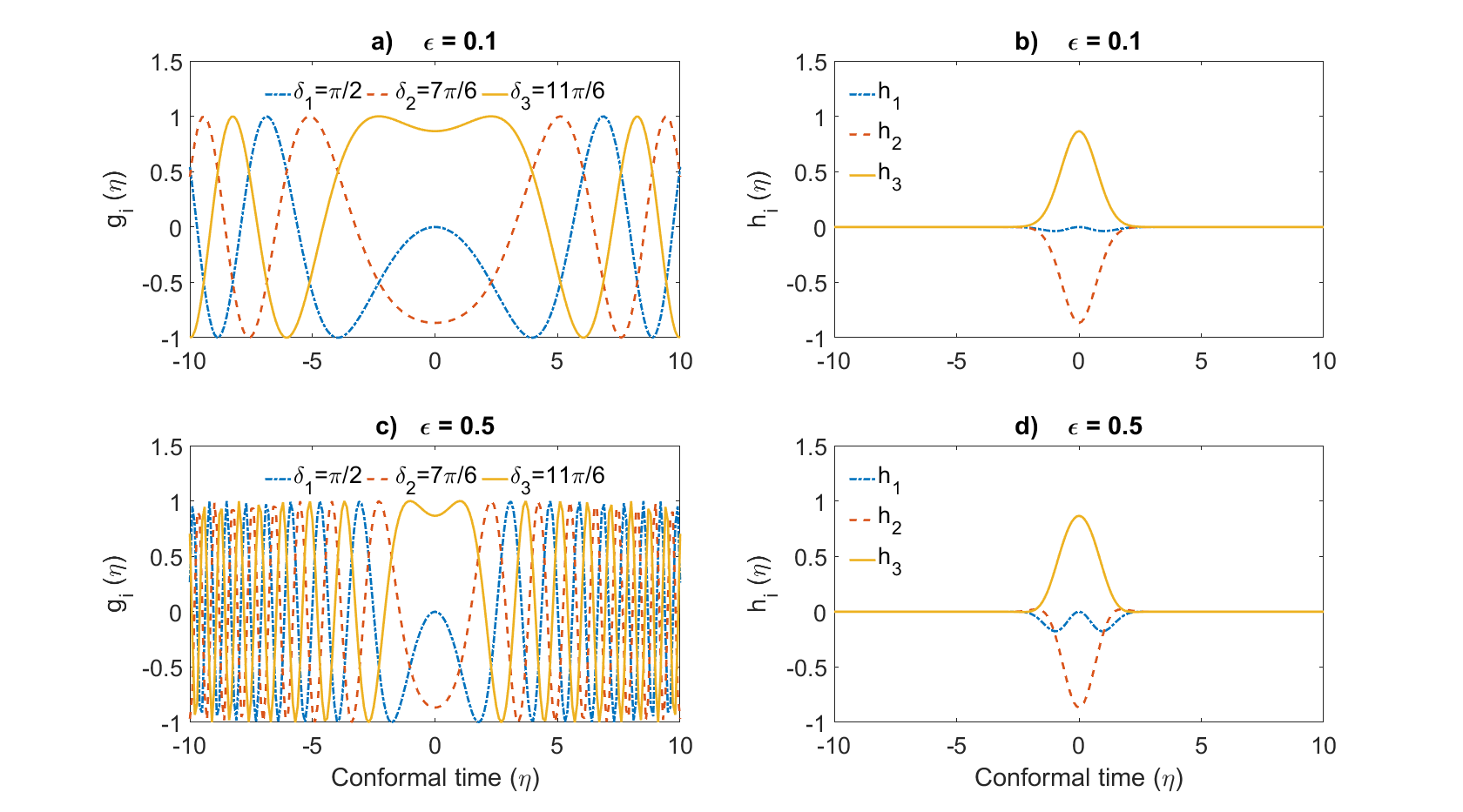}
\caption{Left: oscillatory part of the anisotropic perturbation, $g_i(\eta)$ for a) $\epsilon=0.1$ and c) $\epsilon=0.5$. Right: anisotropic perturbation $h_i(\eta)$ for b) $\epsilon=0.1$ and d) $\epsilon=0.5$ }
\label{fig:h1h2h3}
\end{figure}

Let us now look into a specific model of an anisotropic universe filled with a matter field.
The conformal symmetry breaking that leads to particles production can arise by the departure of the background space-time from conformal flatness. 
The metric can be simply defined by perturbing a FRW metric
\b\label{metric}
ds^2=a^2(\eta)\left\{d\eta^2-\left[1+h_i(\eta)\right]\left(dx^i\right)^2\right\} \,
\e
with $i=1,2,3$.
The perturbation is considered to be small such that $\max_\eta |h_i(\eta)|\ll 1$. This is an example of Bianchi type I metric with weak anisotropy. Imposing $\sum_{i=1}^3 h_i(\eta)=0$ to the perturbation simplifies the form of Dirac equation. One possible choice that satisfies this condition is \cite{BD01}
\b\label{pert-func}
h_i(\eta)=e^{-\rho\,\eta^2}g_i(\eta) \,\;\;\;\;\;\; \delta_i=\frac{\pi}{2},\frac{7\pi}{6},\frac{11\pi}{6},
\e
where $g_i(\eta)=\cos{(\epsilon\,\eta^2+\delta_i)}$ is the oscillatory part of the anisotropic perturbation. This model describes attenuation of anisotropy in the Robertson-Walker universe. Figures \ref{fig:h1h2h3}a) and \ref{fig:h1h2h3}c) display the role of $\epsilon$ as a frequency of oscillation in $g_i(\eta)$, while figures \ref{fig:h1h2h3}b) and \ref{fig:h1h2h3}d) illustrate how the anisotropic perturbation attenuates on the sides of the conformal time $\eta=0$. As for the isotropic universe, the metric $(\ref{metric})$ is homogeneous and, because of this, the the field modes decouple in the anisotropic case too.
Let us now consider a scale factor which allows us to perform analytical calculations and provide explicit expressions for the isotropic and anisotropic contributions to the Bogolubov coefficients. A suitable choice is the following \cite{Bir02}
\b\label{sc-fac}
a^2(\eta)=1-A\,e^{-\rho^2\,\eta^2}\,
\e
with $A$ and $\rho$ real positive constants.  It represents a contracting universe which bounces back at $\eta=0$ and expands out again.

%%%%%%%%%%%%%%%%%%%%%%%%%%%%%%%%%
\subsection{Scalar field}

A massive bosonic field $\phi(\vec{x},t)$ of mass $m$ obeys the Klein-Gordon equation
\b\label{KG}
\left(\Box_g+m^2+\xi R(\eta)\right)\phi(\vec x,\eta)=0 \,,
\e
where $\Box_g$ is the D'Alambertian generalized to the metric $g$
\b
\Box_g=\frac{1}{\sqrt{-g}}\partial_{\mu}(\sqrt{-g}g^{\mu\nu}\partial_{\nu}),
\e
and the factor $\xi$ represents the coupling constant of the field with the Ricci scalar curvature $R$. Since the spacetime is homogeneous, the solution of the field equation \eqref{KG} can be separated as
\b\label{eig-sol}
\phi_{\vec k}(\vec x,\eta)=\frac{1}{(2\pi)^{3/2}}\,\frac{1}{a(\eta)}e^{i\vec k\cdot\vec x}\chi_{\vec k}(\eta) \,.
\e
The function of the time parameter $\chi_{\vec k}(\eta)$ satisfies
\b\label{time-eq}
\ddot\chi_{\vec k}(\eta)+[\omega_{\vec k}^2-V_{\vec k}(\eta)]\chi_{\vec k}(\eta)=0 \,,
\e
where
$\omega^2=k^2+m^2a^2(\infty)$ and
\b\label{V-func}
V_{\vec k}(\eta)=\sum_ih_i(\eta)k_i^2+m^2\left[a^2(\infty)-a^2(\eta)\right]-\left(\xi-\frac{1}{6}\right) R(\eta)a^2(\eta)\,.
\e
The function $V_{\vec k}(\eta)$ is the sum of three contributions
\begin{subequations}\label{Vs}
\begin{align}
&V_{\vec k}^{(mass)}(\eta)=m^2\left[a^2(\infty)-a^2(\eta)\right]\,, \\
&V_{\vec k}^{(curv)}(\eta)=-\left(\xi-\frac{1}{6}\right)  R(\eta)a^2(\eta) \,,\\
&V_{\vec k}^{(aniso)}(\eta)=\sum_ih_i(\eta)k_i^2 \,,
\end{align}
\end{subequations}
coming from mass, curvature coupling and anisotropy respectively. Two regimes are of special interest in cosmology: one is when $\xi=0$, called weak coupling, and the other is when $\xi=1/6$, called conformal coupling, for which $V^{(curv)}=0$. For simplicity, and for the purpose of comparison with the Dirac field, we will consider only the conformal coupling scenario in the following.

We can think to the problem as in scattering theory, which means assuming the interaction of the gravitational field with the matter field to be zero in the early past and in the far future, i.e. $V_k(\pm\infty)=0$. Our choice of the involved functions is such that this condition is matched for every single contribution to the time dependent function $V_k(\eta)$, namely $\lim_{\eta\to\pm\infty}V_{\vec k}^{(\ldots)}(\eta)=0$.
The normalized free-wave solution, propagating from $\eta\to-\infty$, reads
\b\label{free-wave}
\chi^{in}_{\vec k}(\eta)=\frac{1}{\sqrt{2\omega}}e^{-i\omega\eta} \,.
\e
The integral form of the differential equation (\ref{time-eq}) then becomes
\b\label{sol}
\chi_{\vec k}(\eta)=\chi^{in}_{\vec k}(\eta)+\frac{1}{\omega}\int_{-\infty}^{\eta}\,d\eta_1\,V_{\vec k}(\eta_1)\sin\left(\omega(\eta-\eta_1)\right)\chi_{\vec k}(\eta_1) \,.
\e
In late time regions $(\eta\to\infty)$, again, we have a free propagating wave
\b
\chi^{out}_{\vec k}(\eta)=\alpha_{\vec k}\chi^{in}_{\vec k}(\eta)+\beta_{\vec k}\chi_{\vec k}^{in*}(\eta) \,,
\e
where the Bogolubov coefficients $\alpha$ and $\beta$ result as
\begin{subequations}\label{bog-coef}
\begin{align}
\alpha_{\vec k}&=1+i\int_{-\infty}^{\infty}\chi_{\vec k}^{in*}(\eta)V_{\vec k}(\eta)\chi_{\vec k}(\eta)\,d\eta \,, \\
\beta_{\vec k}&=-i\int_{-\infty}^{\infty}\chi_{\vec k}^{in}(\eta)V_{\vec k}(\eta)\chi_{\vec k}(\eta)\,d\eta \,.
\end{align}
\end{subequations}
Computing the Wronskian of the solutions to the differential equation (\ref{sol}) results in Eq. (\ref{norm1}).
To solve Eq. (\ref{sol}) we resort to an iterative procedure. The lowest order gives
\b\label{appr}
\chi_{\vec k}(\eta)=\chi_{\vec k}^{in}(\eta)\,,
\e
and
\begin{subequations}\label{bog-coef-appr}
\begin{align}
\alpha_{\vec k}&=1+\frac{i}{2\omega}\int_{-\infty}^{\infty}V_{\vec k}(\eta)\,d\eta \,,\\
\beta_{\vec k}&=-\frac{i}{2\omega}\int_{-\infty}^{\infty}e^{-2i\omega\eta}V_{\vec k}(\eta)\,d\eta \,.
\end{align}
\end{subequations}
Inserting here the functions (\ref{Vs}) with the explicit form of $h_j(\eta)$ and $a(\eta)$ we can write the Bogolubov coefficients as
\begin{subequations}\label{bog-coef3}
\begin{align}
\alpha_{\vec k}(\eta)&=1+\alpha^{(iso)}_{\vec k}+\alpha^{(aniso)}_{\vec k} \,,  \\
\beta_{\vec k}(\eta)&=\beta^{(iso)}_{\vec k}+\beta^{(aniso)}_{\vec k}  \,.
\end{align}
\end{subequations}
The Bogolubov coefficients (\ref{bog-coef3}) are reported explicitly
in Appendix \ref{App1}.
They have a independent contribution from the mass $m$ and from the anisotropy $h_i$, which are responsible for particles creation and for entanglement generation.
The contribution from the mass turns out to be fully isotropic.

We remark that caution must be used with the approximate solution of (\ref{sol}), and that the Wronskian condition (\ref{norm1}) can be employed as a test for its validity \cite{Bir01}.

The particle reduced density operator takes the same form of \eqref{redscal}
\begin{align}\label{redani1}
\varrho_{\vec k}^{(out)}={\rm Tr}_{-\vec k}\left(\varrho_{\vec k,-\vec k}^{(out)}\right)
=\left(1-\left|\frac{\beta_{\vec k}^*}{\alpha_{\vec k}}\right|^2\right)\sum_{n=0}^{\infty}
\left|\frac{\beta_{\vec k}^*}{\alpha_{\vec k}}\right|^{2n} |n\rangle_{\vec k}\langle n|^{out},
\end{align}
where this time $\alpha_{\vec k}$, $\beta_{\vec k}$ are given by \eqref{bog-coef3}.

%%%%%%%%%%%%%%%%%%%%%%%%%

\subsection{Dirac Field}

The matter field dynamics is given by the following Dirac equation
\begin{equation}
\tilde{\gamma}^{\mu}(\partial_{\mu}+\Gamma_{\mu})\psi+m\psi=0 \,,
\label{dir-eq}
\end{equation}
where
$\tilde{\gamma}^{\mu}\Gamma_{\mu}=\frac{3}{2}\frac{\dot a}{a^2}\gamma^0$.
The details of the solution and the explicit form of the Bogolubov coefficients are given in the Appendix \ref{App2}. It is shown there, that with the choice (\ref{pert-func}) and (\ref{sc-fac}) for the perturbative function and the scale factor respectively, the following relations hold
\begin{subequations}\label{abrel}
\begin{align}
& \alpha_{\vec k\,\u\d}=\alpha_{\vec k\,\d\u}=0\,, \\
& \alpha_{\vec k\,\d\d}=\alpha_{\vec k\,\u\u}=\alpha_{\vec k}\,,  \\
&\beta_{\vec k\,\u\u}=-\beta_{\vec k\,\d\d}\,, \\
&\beta_{\vec k\,\u\d}=-\beta_{\vec k\,\d\u}^* \,.
\end{align}
\end{subequations}
Consequently, the operator $a_{\vec k\,\sigma}^{in}$, defined in $(\ref{dir-bog})$, simplifies to
 \begin{equation}\label{dir-bog1}
a_{\vec k\,\sigma}^{in}=\alpha(\vec k)^*a_{\vec k\,\sigma'}^{out}+ \sum_{\sigma'}\beta^*_{\vec k\sigma\sigma'}b^{\dagger\,out}_{-\vec k\,\sigma'} \,,
\end{equation}
and the Wronskian condition becomes
\b\label{norm2}
|\alpha_{\vec k}|^2+|\beta_{\vec k\,\uparrow\uparrow}|^2+|\beta_{\vec k\,\uparrow\downarrow}|^2=1 \,,
\e
which should also be used to define the range of validity of approximate solution.

As for the scalar field it is possible to explicitly separate the isotropic from the anisotropic part
\begin{subequations}
\begin{align}
&\alpha_{\vec k}=1+\alpha_{\vec k}^{iso}+\alpha_{\vec k}^{aniso}\,,  \\
&\beta_{\vec k\,\uparrow\uparrow}=\beta^{iso}_{\vec k\,\uparrow\uparrow}+\beta_{\vec k\,\uparrow\uparrow}^{aniso}\,,  \\
&\beta_{\vec k\,\uparrow\downarrow}=\beta_{\vec k\,\uparrow\downarrow}^{iso}
+\beta_{\vec k\,\uparrow\downarrow}^{aniso}   \,.
\end{align}
\end{subequations}
The isotropic contribution is dependent upon the mass and is equal to zero when $m=0$, while the anisotropic part is clearly dependent upon the functions $h_j$.

The off diagonal elements in the density matrix (\ref{reducedrho}) are then zero, and the reduced density matrix reads
\begin{align}
\varrho_{\vec k}^{(out)}= {\rm Tr}_{-\vec k}\left(\varrho_{\vec k,-\vec k}^{(out)}\right)=
\frac{|\beta_{\vec k\,\u\u}|^2}{|\beta_{\vec k\,\d\d}|^2}
\Big[&|\alpha_{\vec k}|^4 |0_{\vec k}\ra\la 0_{\vec k}|
+|\alpha_{\vec k}|^2( |\beta_{\vec k\,\u\d}|^2+|\beta_{\vec k\,\u\u}|^2) (|\u_{\vec k}\ra\la\u_{\vec k}|+ |\d_{\vec k}\ra\la\d_{\vec k}|)\nonumber\\
&+ (|\beta_{\vec k\,\u\d}|^2 +|\beta_{\vec k\,\u\u}|^2)^2 |\u\d_{\vec k}\ra\la\u\d_{\vec k}| \Big] \,.
\label{reducedrho2}
\end{align}

%%%%%%%%%%%%%%%%%%%%%%%%%%%%%%%%%%%%%%%%%%%%%%%%%%%%%%%%%%%%%%%%

\section{Results about entanglement}\label{section4}

In this section we compare and point out differences between the entanglement entropy of the scalar and the Dirac field, as well as the isotropic (mass) and the anisotropic contribution to it. The entanglement entropy takes the following form
\begin{align}\label{entr-funct} 
S\left(\vec{k}\right)=-{\rm Tr}\left(\varrho_{\vec k}^{(out)}\log_2 \varrho_{\vec k}^{(out)}\right)
=(1-3 \delta_F) |\alpha_{\vec k}|^2 \log |\alpha_{\vec k}|^2-(1-3 \delta_F) (|\alpha_{\vec k}|^2-1) \log [ (1-2\delta_F) ( |\alpha_{\vec k}|^2-1) ]   \,,
\end{align}
where $\delta_F$ equals $0$ for the scalar and $1$ for Dirac field.
Eq.\eqref{entr-funct} has been derived using \eqref{redani1} for the scalar field 
and \eqref{reducedrho2} (together with \eqref{abrel}) for the Dirac field.
There $\alpha_{\vec k}=1+\alpha_{\vec k}^{iso}+\alpha_{\vec k}^{aniso}$, where
$\alpha_{\vec k}^{iso}$ and $\alpha_{\vec k}^{aniso}$ are explicitly given in Appendix \ref{App1} and Appendix \ref{App2}, for the scalar field and Dirac field respectively.

\begin{figure}[ptb]
\begin{minipage}[h]{0.45\textwidth}
\includegraphics[width=2.5in]{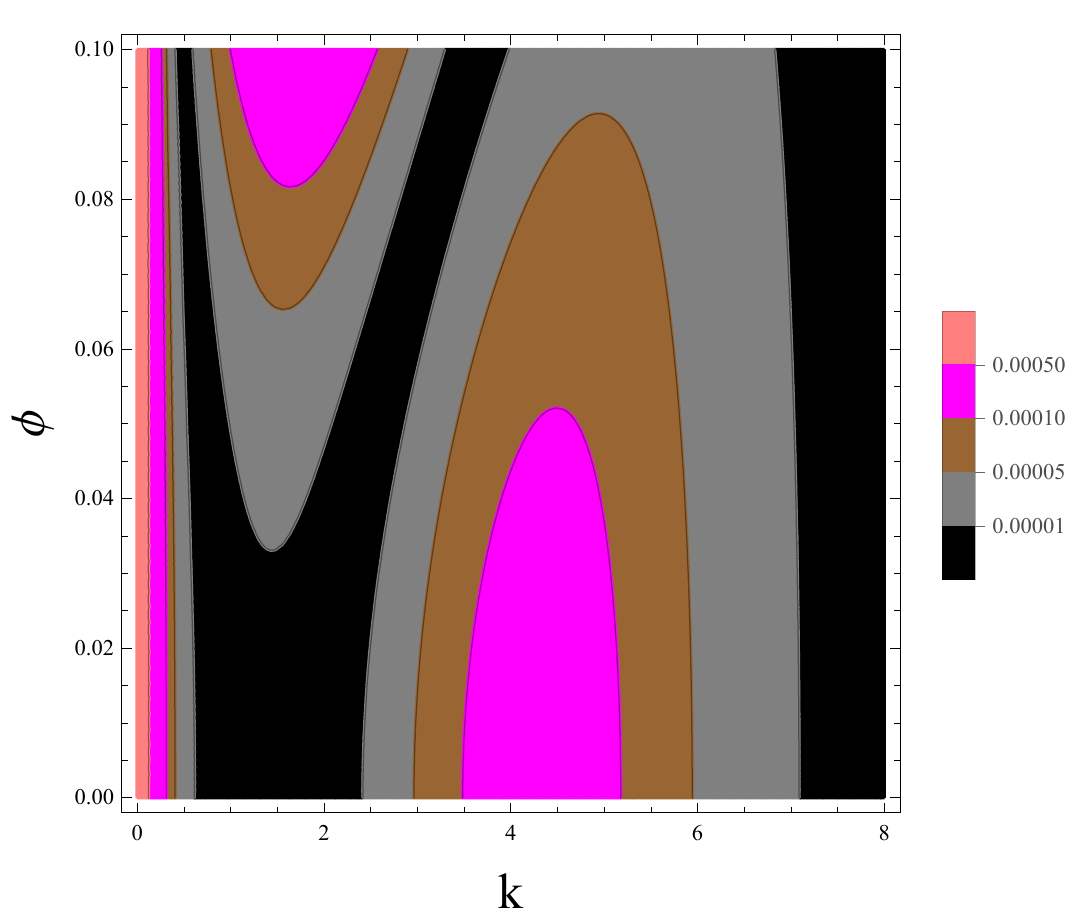}	
\end{minipage}
\begin{minipage}[h]{0.45\textwidth}
\includegraphics[width=2.5in]{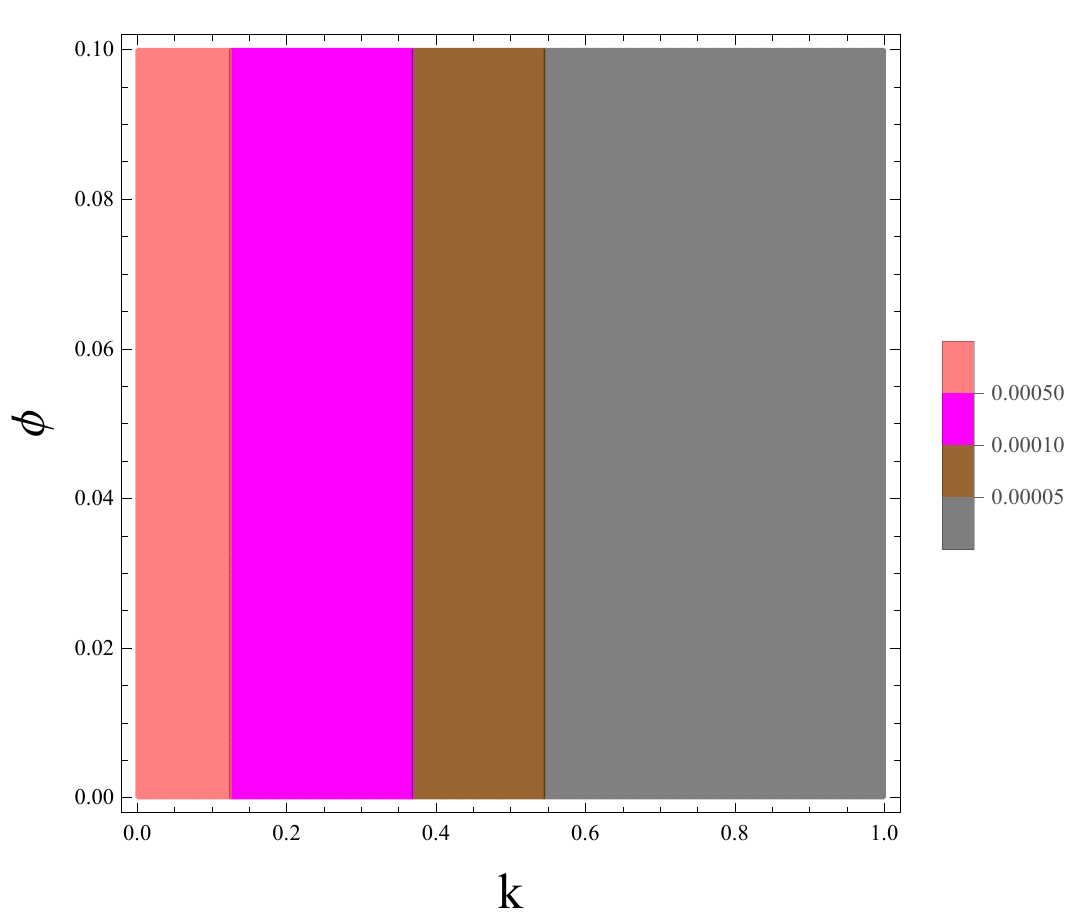}
\end{minipage}
\caption{ Density plot of the total subsystem entropy $S$ (left) and of the isotropic (mass) subsystem entropy (right) of the scalar field vs $k$ and $\phi$.
The values of the parameters are $\rho=10$, $\varepsilon=0.1$, $m=0.1$,
$\theta=\frac{\pi}{2}$.}
\label{fig1}
\end{figure}

\begin{figure}[ptb]	
\begin{minipage}[h]{0.45\textwidth}
\includegraphics[width=2.5in]{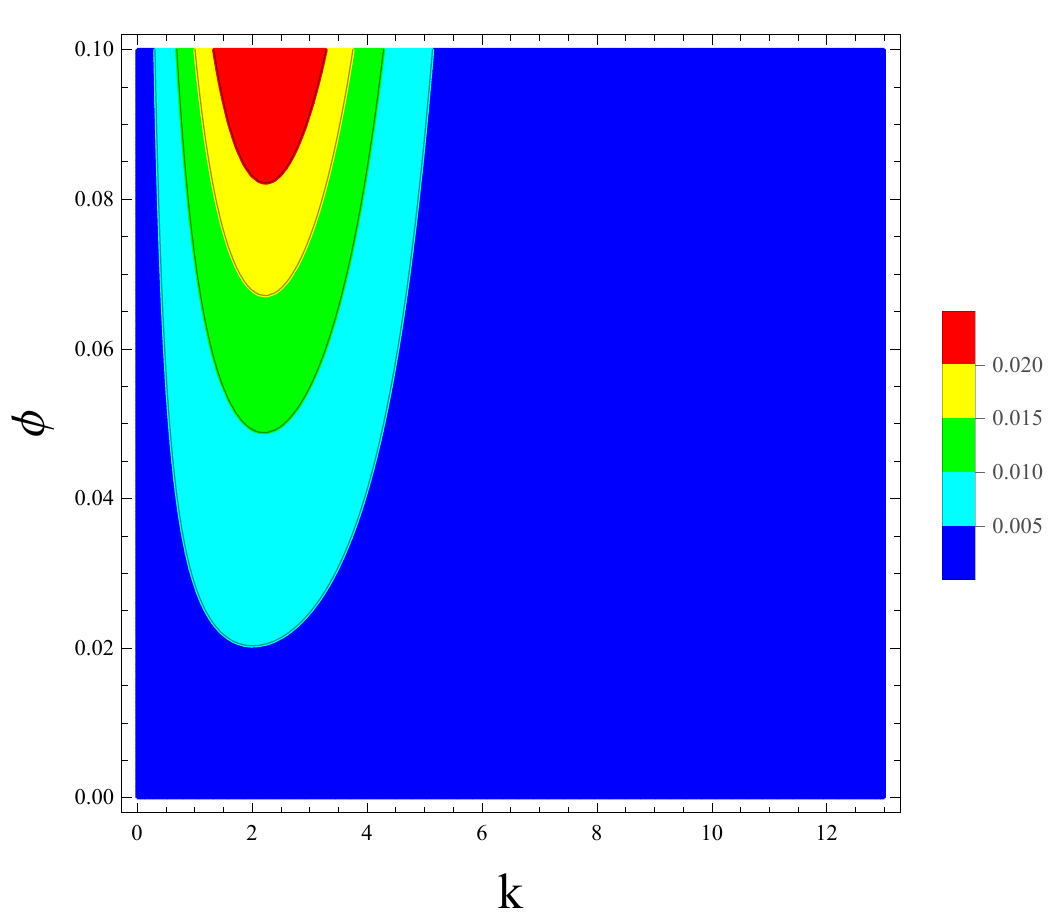}
\end{minipage}
\begin{minipage}[h]{0.45\textwidth}
\includegraphics[width=2.5in]{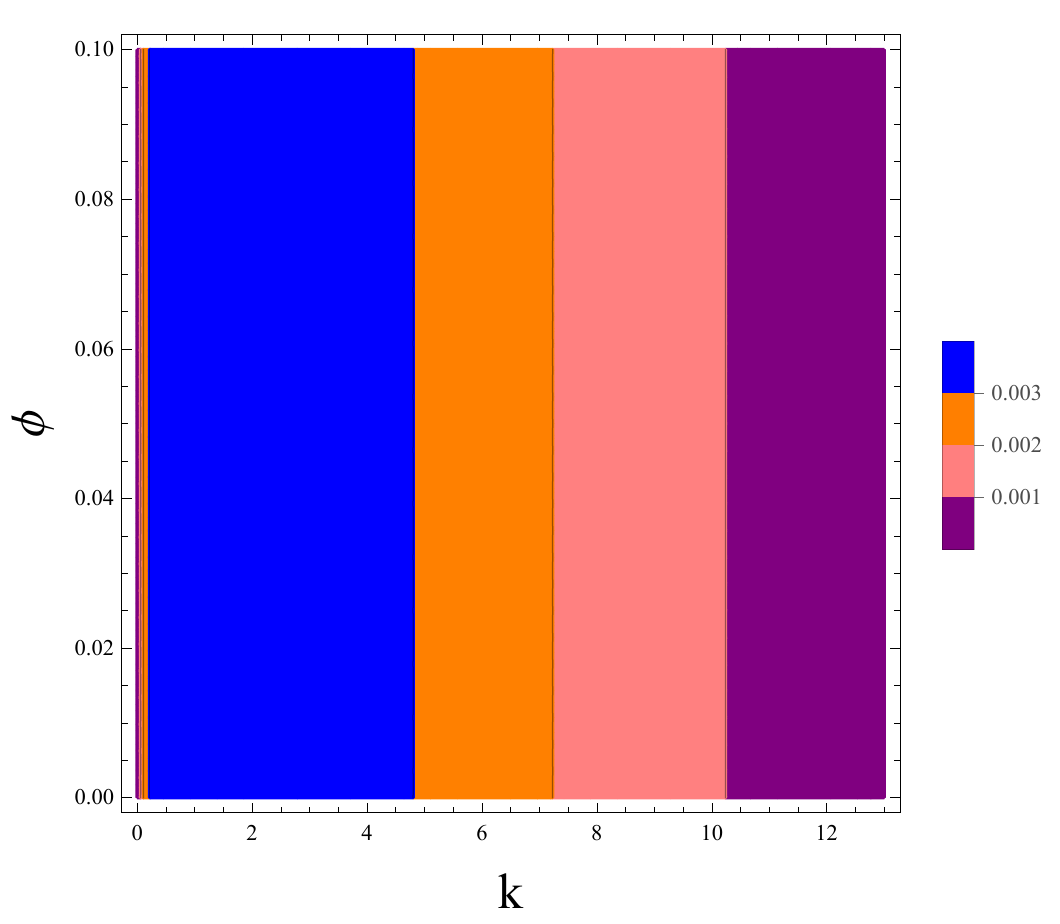}
\end{minipage}
\caption{ Contour plot of the total subsystem entropy $S$ (left) and of the isotropic (mass) subsystem entropy (right) of the Dirac field vs $k$ and $\phi$.
The values of parameters are $\rho=10$, $\varepsilon=0.1$, $m=0.1$ and $\theta=\frac{\pi}{2}$.}
\label{fig2}
\end{figure}

\begin{figure}[ptb]
\begin{minipage}[h]{0.45\textwidth}
\includegraphics[width=2.5in]{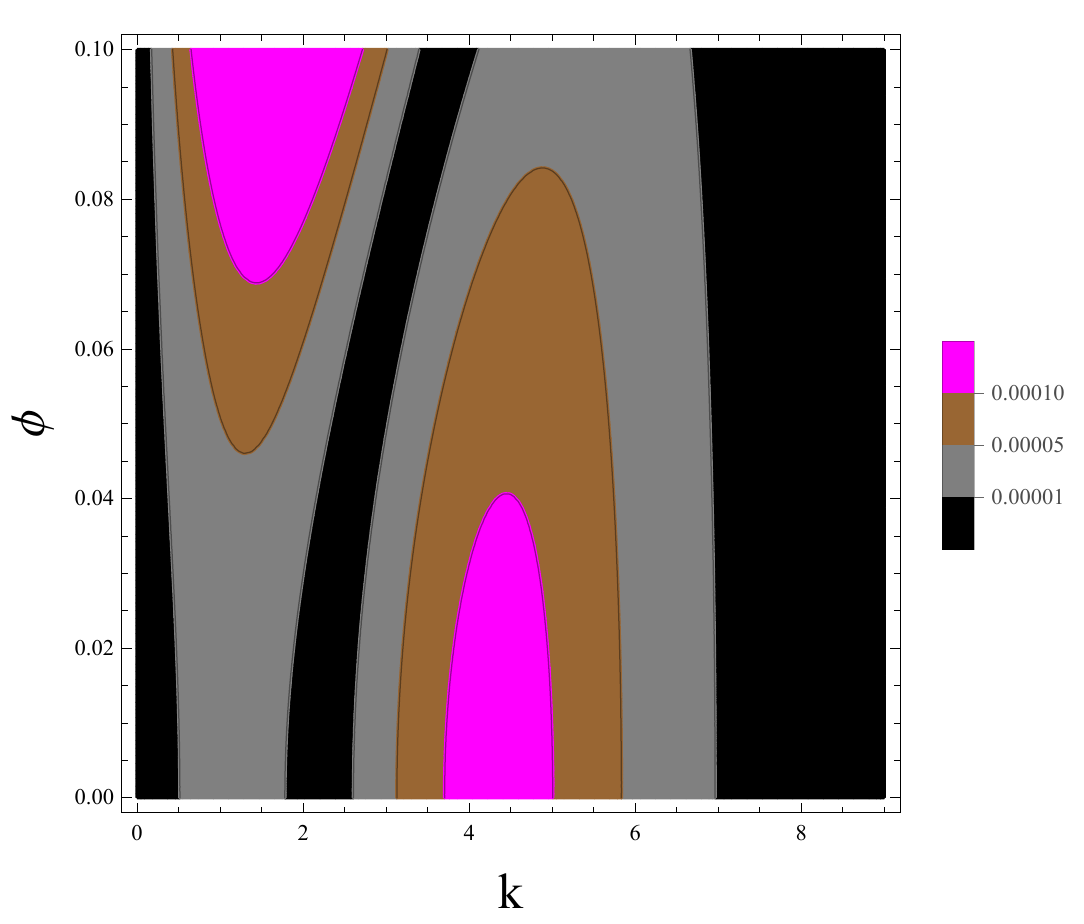}
\end{minipage}
\begin{minipage}[h]{0.45\textwidth}
\includegraphics[width=2.5in]{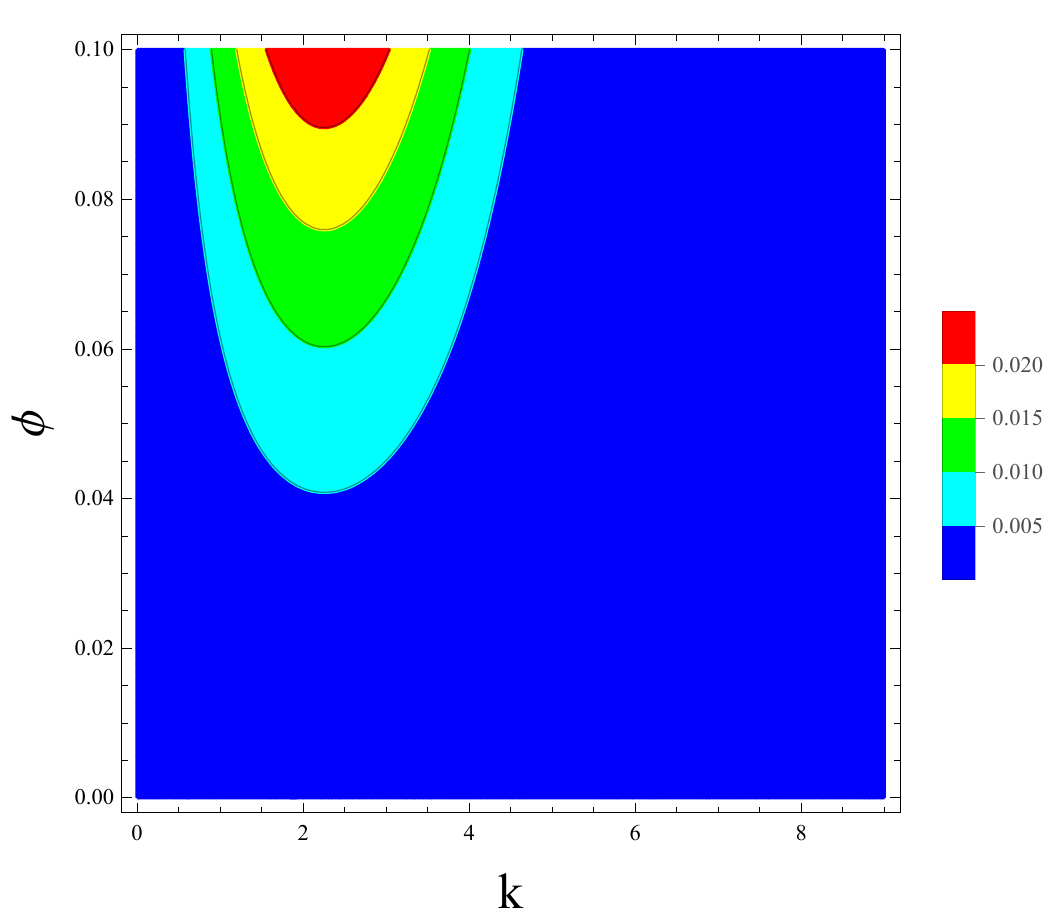}
\end{minipage}
\caption{Density plot of the anisotropic subsystem entropy $S$ of the massless scalar field (left) and of the massless Dirac field (right) vs $k$ and $\phi$.
The values of parameters are $\rho=10$, $\varepsilon=0.1$, $m=0.0$ and
$\theta=\frac{\pi}{2}$.}
\label{fig5}
\end{figure}

In Figures \ref{fig1}-\ref{fig5} we show how the entanglement entropy \eqref{entr-funct} depends on the momentum
\begin{align}
\vec{k}=\left(k\sin\theta\cos\phi,k\sin\theta\sin\phi,k\cos\theta\right).
\end{align}
It is important to note at this point that, while the formula $(\ref{entr-funct})$ for the entanglement entropy is exact, the explicit expression of the Bogulubov coefficients hold approximately only. A resonable way to make sure our plots are meaningful is to accept a deviation from the Wronskian conditions $(\ref{norm1})$, for the scalar field, and $(\ref{norm2})$, for the Dirac field, of one percent at most. The choice of the parameters and the momentum range in the plots are a consequence of this.   
The anisotropic contribution to entanglement is expected to depend upon the direction of the particle momentum.
Then we show each plot as a function of the modulus $k$ of the momentum and the polar angle $\phi$, by fixing the azimuthal angle $\theta$\footnote{The choice of fixing $\theta$, 
instead of $\phi$ in the plots, is motivated by purely aesthetic reasons.}.
Fig. \ref{fig1} and \ref{fig2} are instances of the massive scalar and Dirac field respectively. Actually the left plots in Fig. \ref{fig1} and \ref{fig2} represent the total entropy (which clearly changes by changing 
$\phi$), while the right ones only show the contribution from the mass, which is not sensible to the direction. It is worth remarking the differences between the two fields. In the scalar case we find that anisotropy gives rise to revivals of entanglement entropy after decreasing from the maximum at $k = 0$. For Dirac field the effect is a slight distortion of the non-monotonic profile giving rise to the maximum of entanglement entropy at $k > 0$. Furthermore, the mass contribution to the scalar field entanglement is relevant only at small momenta, where it is responsible for the main contribution to the entropy. This is not the case for the Dirac field.

The role of anisotropy can be better understood from Fig. $(\ref{fig5})$, where the mass is set to zero. In this case both fields have vanishing entanglement at zero momentum.
However they present a not negligible entanglement at $k>0$. This means that anisotropies in the expansion of the universe have an important role in forcing quantum correlations between particles. Such entanglement, depending on the space direction, greatly prevails in the Dirac particles (as can be seen from the different scale of the density plots in Fig. \ref{fig5}).

\vspace{.5cm}

\section{Concluding Remarks}\label{section5}

We have shown that spacetime anisotropies affect the particle-antiparticle quantum correlations. Although we have used a toy model to overcome technical difficulties in solving dynamical equations, the results are enlightening. We found that revivals of entanglement entropy vs momentum appear in the massive scalar field and they also depend on the space direction. For the Dirac field the effect is a slight distortion of the non-monotonic profile giving rise to the maximum of entanglement entropy at $k > 0$.

More interestingly, massless field of both type can only get entangled through anisotropy. This is of particular relevance for massless Dirac particles like neutrino which, being weakly interacting with other fields, may have not completely washed out correlations in their evolution to present time. Detecting cosmic neutrino background \cite{Man,Tro} is a challenging task, but attempts are on the way \cite{Fol}. Furthermore, it is known that anisotropies in the underling metric reflect on temperature fluctuations of the CMB radiation \cite{Muk}. Given that anisotropies are also responsible of massless particles creation and of the entanglement coming along, we can think of the temperature fluctuations as a signature of quantum correlations. On the other way around, we can think to extract information on spacetime parameters from entanglement.    
An operative way has been recently presented in Ref. \cite{steeg09}, where authors have shown that two spacelike separated detectors, placed in satellites and interacting with the same field, become entangled in a way that is sensitive to the spacetime background. Therefore, we envisage the possibility of applying these same methods to reveal anisotropic effects. A broad discussion regarding possible experiments to realize with satellites can be found in Ref. \cite{RID}.

We thus believe that the presented study, not only can integrate those aimed at estimating cosmological parameters by means of entanglement \cite{FMMM}, but it can also pave the way to high resolution `tomography' of universe to highlight its clumsiness. 
This should foresee also investigations of non-homogeneity besides anisotropy of spacetime. 
Along this line one could consider the more realistic metric (in Newtonian gauge) 
\b
ds^2=a(\eta)^2\left[(1+2\Phi(\eta,\vec x))d\eta^2-(1-2\Phi(\eta,\vec x))\delta_{ij}dx^idx^j\right]  \,,
\e
with a function $\Phi(\eta,\vec x))$ depending on the space coordinates $\vec x$, besides time $\eta$.
As consequence the field modes would not decouple and the particles would not be simply created as entangled pairs with opposite momenta, but multimode entanglement will arise. 
However a completely different approach should be taken for its quantitative investigation.  
Finally, we may notice that the study of entanglement in anisotropic spacetime can shed light on the degradation of information due to universe evolution \cite{MPW14}. This in fact should occur in the same directions as anisotropies emerge in the universe. These coefficients depend on the cosmological parameters defining the metric $(\ref{metric})$.

%%%%%%%%%%%%%%%%%%%%%%%%%%%%%%%%%%%%%%%%%%%%%%%%%%%

\acknowledgments
This work has been supported by FQXi under the programme ``Physics of Observer 2016".
R. P. thanks the National Science Centre, Sonata BIS Grant No. DEC-2012/07/E/ST2/01402 for financial support.
%%%%%%%%%%%%%%%%%%%%%%%%%%%%%%%%%%%%%%%%%%%%%%%%%%%
%%%%%%%%%%%%%%%%%%%%%%%%%%%%%%%%%%%%%%%%%%%%%%%%%%%%%%%%%%%%%%%%%%%%%%%%%%%%%%%%%%%%%%%%%%%%%%%%%%%%%%%%%%%%%%%%%%%%%

\appendix

\section{Derivation of the Bogolubov coefficients for the scalar field}\label{App1}
Approximate expressions for the Bogolubov coefficients of the scalar field can be analytically computed inserting Eq. \eqref{Vs} into \eqref{bog-coef-appr}). With the scale factor $a(\eta)$ of \eqref{sc-fac}) and and the perturbations $h_i(\eta)$ of \eqref{pert-func}) we get integrals of Gaussian-like functions. They can be easily computed to give
\begin{equation}
\begin{aligned}\label{sca-coef}
&\alpha_{\vec k}^{iso}=1+\frac{i m^2\,A\sqrt{\pi}}{2\omega\rho}\,,  \\
&\beta^{iso}_{\vec k}=-\frac{i m^2\,A\sqrt{\pi}}{2\omega\rho}e^{-\omega^2/\rho^2}\,,
\end{aligned}
\end{equation}
and 
\begin{equation}
\begin{aligned}
\label{scalar-Bog-coef}
&\alpha^{(aniso)}_{\vec k}= \frac{i\sqrt{\pi}}{2\omega}\sum_{j=1}^3 k_j^2 \,{\rm Re}\left\{ \frac{e^{-i\delta_j}}{\sqrt{\rho+i\varepsilon}} \right\}\,,\\
&\beta^{(aniso)}_{\vec k}=- \frac{i\sqrt{\pi}}{2\omega}\sum_{j=1}^3 k_j^2 \,{\rm Re}\left\{ \frac{e^{-i\delta_j}\,e^{-\omega^2/(\rho+i\varepsilon)}}{\sqrt{\rho+i\varepsilon}} \right\}\,,
\end{aligned}
\end{equation}
where $\delta_i$ is the phase of \eqref{pert-func}.

%%%%%%%%%%%%%%%%%%%%%%%%%%%%%%%%%%%%%

\section{Derivation of the Bogolubov coefficients for the Dirac field}\label{App2}

The procedure to compute the Bogolubov coefficients for the Dirac field 
is rather involved and report here only the main steps.
Following \cite{Lot02} we insert $\psi=a^{-3/2}\phi$ in (\ref{dir-eq}) to get
\begin{equation}
\gamma^{\mu}\partial_{\mu}\phi+ma(\eta)\phi=\frac{1}{2}\sum_{j=1}^3h_j\gamma^m\partial_{j}\phi \,.
\end{equation}
Making the ansatz $\phi=\gamma^{\nu}\partial_{\nu}\varphi-\frac{1}{2}\sum^3_{j=1} h_j\gamma^j\partial_{j}\varphi-ma(\eta)\varphi$, and taking only first orders terms in $h_i$, we get the so-called iterated Dirac equation
\begin{equation}\label{ite-dirac-eq}
\eta^{\mu\nu}\partial_{\mu}\partial_{\nu}\varphi-m\dot a(\eta)\gamma^0\varphi-m^2a^2\varphi=\sum_j(h_j\eta^{ij}\partial_{j}^2+\frac{1}{2}\dot h_j\gamma^j\gamma^0\partial_j) \,.
\end{equation}
Now we look for solutions in the following form
\begin{equation}\label{iter-eq-sol}
\varphi=f^{(\varepsilon,\sigma)}(\eta)\exp(i \vec k\cdot\vec x )\,,
\end{equation}
where
\b
f^{(-,+)}=\left(
\begin{array}{c}
\exp(-i\omega\eta)+F_1(\eta)^{-,+}  \\
F_2(\eta)^{-,+}  \\
F_3(\eta)^{-,+}  \\
F_4(\eta)^{-,+}
\end{array}
\right),
\qquad\qquad
f^{(-,-)}=\left(
\begin{array}{c}
F_1(\eta)^{-,+}  \\
\exp(-i\omega\eta)+F_2(\eta)^{-,+}  \\
F_3(\eta)^{-,+}  \\
F_4(\eta)^{-,+}
\end{array}
\right),
\e
\b
f^{(+,+)}=\left(
\begin{array}{c}
F_1(\eta)^{-,+}  \\
F_2(\eta)^{-,+}  \\
\exp(i\omega\eta)+F_3(\eta)^{-,+}  \\
F_4(\eta)^{-,+}
\end{array}
\right),
\qquad\qquad
f^{(+,-)}=\left(
\begin{array}{c}
F_1(\eta)^{-,+}  \\
F_2(\eta)^{-,+}  \\
F_3(\eta)^{-,+}  \\
\exp(i\omega\eta)+F_4(\eta)^{-,+}
\end{array}
\right).
\e
For $\eta\to\pm\infty$ such solutions must reduce to the standard exponential solutions familiar in the Minkowski spacetime 
\b
f^{-,\sigma}=u_{\sigma}\,e^{-i\omega\eta} \qquad\qquad \text{and} \qquad\qquad  f^{+,\sigma}=v_{\sigma}\,e^{i\omega\eta}  \,.
\e
The $Fs$ are small, that means perturbative quantities, and satisfy the following system of differential equations
\begin{equation}
\ddot F^{(\pm,\sigma)}_i(\eta)+\omega^2F^{(\pm,\sigma)}_i(\eta)=V_i^{(\pm,\sigma)}(\eta)\exp(i\pm\omega\eta) \,,
\end{equation}
where
\begin{align*}
V_1^{(-,+)}&=V_2^{(-,-)}=V_3^{(+,+)*}=V_4^{(+,-)*}=m^2-m^2a^2+im\dot{a}(\eta)+\sum h_i k_i^2 \,,  \\
V_2^{(-,+)}&=V_1^{(-,-)}=V_4^{(+,+)}=V_3^{(+,-)}=0  \,,  \\
V_3^{(-,+)}&=-V_4^{(-,-)}=V_1^{(+,+)}=-V_2^{(+,-)}=\frac{1}{2}\, i\,k_3 \,\dot h_3       \,,  \\
V_4^{(-,+)}&=-V_3^{(-,-)*}=V_2^{(+,+)}=-V_1^{(+,-)*}=\frac{1}{2}i(k_1\, \dot h_1+i\, k_2\,\dot h_2)   \,.
\end{align*}
As we did for the scalar field the differential equation can be transformed into an integral equation and we can define two different sets of solutions imposing two different boundary conditions. The set of solutions $F^{(\pm,\sigma)}_{in}(\eta)$ such that  $F^{(\pm,\sigma)}_{i\,,in}(-\infty)=0$ is
\begin{equation}\label{Fs-in}
F^{(\pm,\sigma)}_{i\,,in}(\eta)=a_i^{(\pm,\sigma)}e^{i\pm\omega\eta}+b_i^{(\pm,\sigma)}e^{-i\pm\omega\eta} \,,
\end{equation}
with
\begin{align}
\begin{split}\label{abs}
a_i^{(\pm,\sigma)}&=\mp\frac{i}{2\omega}\int^{+\eta}_{-\infty} V^{(\pm,\sigma)}_i(\eta) \, d\eta    \,,    \\
b_i^{(\pm,\sigma)}&=\pm\frac{i}{2\omega}\int^{+\eta}_{-\infty} V^{(\pm,\sigma)}_i(\eta)e^{\pm\,2i\omega\eta} \, d\eta \,.
\end{split}
\end{align}
The set of solutions $F^{(\pm,\sigma)}_{i\,,out}(\eta)$ such that  $F^{(\pm,\sigma)}_{i\,,out}(+\infty)=0$ reads
\begin{equation}\label{Fs-out}
F^{(\pm,\sigma)}_{i\,,in}(\eta)=-a_i^{(\pm,\sigma)}e^{i\pm\omega\eta}-b_i^{(\pm,\sigma)}e^{-i\pm\omega\eta} \,.
\end{equation}
Then the solution of the iterated Dirac equation is found inserting \eqref{Fs-in} and \eqref{Fs-out} into \eqref{iter-eq-sol}. From this, we can finally write down the solution of the Dirac equation (\ref{dir-eq}) as
\b
\psi_{\vec k,\sigma}(\vec x,\eta)^{(\pm\,in/out)}=\psi_{\vec k,\sigma}(\eta)^{(\pm\,in/out)}\,e^{\pm i\,\vec k\cdot\vec x} \,.
\e
Next relating the $in$ and $out$ regions through the Bogolubov transformation $(\ref{psi-in-out})$, we can compute the Bogolubov coefficients through the Dirac scalar product $(\ref{dir-bog-coef})$. Defining $k_{\pm}=k_1\pm i\,k_2$
The final result is 
\begin{align}
\begin{split}
\alpha_{\vec k\d\d}&=1+a_2^{(-,-)}-\frac{k_{+}}{\omega+m}\,a_3^{(-,-)}+\frac{k_{3}}{\omega+m}\,a_4^{(-,-)} \,,  \\
\alpha_{\vec k \u\u}&=1+a_1^{(-,+)}-\frac{k_{-}}{\omega+m}\,a_4^{(-,+)}-\frac{k_{3}}{\omega+m}\,a_3^{(-,+)}  \,, \\
\alpha_{\vec k \d\u}&=-\frac{1}{\omega+m}[k_3\,a_3^{(-,-)}+k_-\,a_4^{(-,-)}]  \,, \\
\alpha_{\vec k \u\d}&=-\frac{1}{\omega+m}[k_+\,a_3^{(-,+)}-k_3\,a_4^{(-,+)}] \,,
\end{split}
\end{align}
and
\begin{align}
\begin{split}
\beta_{\vec k\,\d\d}&=b_4^{(-,-)}-\frac{k_{3}}{k^2}(\omega-m)\,b_2^{(-,-)} \,,  \\
\beta_{\vec k\,\d\u}&=b_3^{(-,-)}+\frac{k_{-}}{k^2}(\omega-m)\,b_2^{(-,-)} \,,  \\
\beta_{\vec k\,\u\u}&=b_3^{(-,+)}+\frac{k_{3}}{k^2}(\omega-m)\,b_1^{(-,+)} \,,  \\
\beta_{\vec k\,\u\d}&=b_4^{(-,+)}+\frac{k_{+}}{k^2}(\omega-m)\,b_1^{(-,+)} \,,
\end{split}
\end{align}
where $a_i^{(\varepsilon,\sigma)}$ and $b_i^{(\varepsilon,\sigma)}$ are defined in \eqref{abs}. 
Explicitly we get
\begin{eqnarray}\label{Dir-bog-coeff}
\alpha_{\vec k}&=&1+\frac{im^2\,A\sqrt{\pi}}{2\omega\rho}+\frac{i}{2\omega}\sum k_j^2 \Re\{ \sqrt{\frac{\pi}{\rho-i\epsilon}}\,e^{i\delta_j} \}\,,  \nonumber\\
&=&\alpha_{\vec k}^{iso}+\alpha_{\vec k}^{aniso}   \\
\beta_{\vec k\,\uparrow\uparrow}&=&-\frac{im^2\,A\sqrt{\pi}}{2\omega\rho}\frac{k_3}{k^2}(\omega-m)\left[\text{exp}(-\frac{\omega^2}{\rho^2})+\frac{\rho^2m\,A}{2\omega}\int_{-\infty}^{\infty}d\eta\,\frac{\eta 
\,e^{-2i\omega\eta}}{\sqrt{1-A\,\text{exp}(-\rho^2\eta^2)}}\right]   \nonumber  \\
&+&\frac{i\,k_3}{2}\Re\{\sqrt{\frac{\pi}{\rho-i\epsilon}}\,\text{exp}(-\frac{\omega^2}{\rho-i\epsilon}+i\frac{\pi}{2}+i\frac{4\pi}{3})\}  \nonumber\\
&-&\frac{ik_3}{2\omega k^2}(\omega-m)\sum k_i^2\Re\big\{\sqrt{\frac{\pi}{\rho-i\epsilon}}\,\text{exp}(-\frac{\omega^2}{\rho-i\epsilon}+i\delta_i)\}\,, \nonumber\\
&=&\beta^{iso}_{\vec k\,\uparrow\uparrow}+\beta_{\vec k\,\uparrow\uparrow}^{aniso} \\
\beta_{\vec k\,\uparrow\downarrow}&=&-\frac{im^2\,A\sqrt{\pi}}{2\omega\rho}\frac{k_+}{k^2}(\omega-m)\left[\text{exp}(-\frac{\omega^2}{\rho^2})+\frac{\rho^2m\,A}{2\omega}\int_{-\infty}^{\infty}d\eta\,\frac{\eta\,e^{-2i\omega\eta}}{\sqrt{1-A\,\text{exp}(-\rho^2\eta^2)}}\right]\nonumber \\
&+&\frac{i}{2}\Big[k_1\Re\big\{\sqrt{\frac{\pi}{\rho-i\epsilon}}\,\text{exp}(-\frac{\omega^2}{\rho-i\epsilon}+i\frac{\pi}{2})\big\}+ik_2\Re\big\{\sqrt{\frac{\pi}{\rho-i\epsilon}}\,\text{exp}(-\frac{\omega^2}{\rho-i\epsilon}+i\frac{\pi}{2}+i\frac{2\pi}{3})\big\}\Big] \nonumber  \\
&-&\frac{ik_+}{2\omega k^2}(\omega-m)\sum k_i^2\Re\big\{\sqrt{\frac{\pi}{\rho-i\epsilon}}\,\text{exp}(-\frac{\omega^2}{\rho-i\epsilon}+i\delta_i)\}\nonumber\\
&=&\beta_{\vec k\,\uparrow\downarrow}^{iso}+\beta_{\vec k\,\uparrow\downarrow}^{aniso}  \,.
\end{eqnarray}

%%%%%%%%%%%%%%%%%%%%%%%%%%%%%%%%%%%%%%%%%%%%%%%%%%%%%%%%%%%%%%%%%%%%%%%%%%%%%%%%%%%%%%%%%%%%%%%%%%%%%%%%%%%%%%%%%%%%%%%%%%%%%%%%%%%%%%%%%%%%%%%%%%%%%%%%%%%%%%%%%%%%%%%%

\end{document}